# Experimental Study of Granular Clogging in a Two-Dimensional Hopper

Shuyang Zhang,[1] Zhikun Zeng,[1] Houfei Yuan,[1] Zhifeng Li,[4] and Yujie Wang[1,2,3,*]

[1]School of Physics and Astronomy, Shanghai Jiao Tong University, Shanghai 200240, China
[2]State Key Laboratory of Geohazard Prevention and Geoenvironment Protection, Chengdu University of Technology, Chengdu 610059, China
[3]Department of Physics, College of Mathematics and Physics, Chengdu University of Technology, Chengdu 610059, China
[4]School of Mathematics and Physics, University of South China, Hengyang 421001, China

We experimentally investigate the clogging process of granular materials in a two-dimensional hopper flow, and present a self-consistent physical mechanism of clogging based on precursory chain structures. We found that these chain structures follow a specific modified restricted random walk, and clogging occurs when they are mechanically stable enough to withstand the flow fluctuations, resulting in the formation of an arch at the outlet. We introduce a simple model that can explain the arch-forming probability by incorporating an analytical expression for chain formation and its transition into an arch. Our results provide insight into the microscopic mechanism of clogging in hopper flow.

The clogging of granular flow in silos or hoppers is ubiquitous and closely related to many industrial processes such as the transportation of granular materials in pipelines [1], as well as daily life events like traffic jams [2,3]. Previous studies, both theoretical and experimental, have investigated the influencing factors on the clogging process, such as outlet size [4,5] and shape [6,7], particle shape [8] and friction [9], external mechanical agitation [10,11] and presence of obstacles above the outlet [12-14]. Nevertheless, the microscopic mechanism for clogging and how it is influenced by these factors is still not yet well understood [15-17].

A hallmark feature of clogging is the formation of an arch near the outlet [18], which can withstand the stress induced by particles above [19]. To *et al*. proposed that the clogging in silo discharge is caused by the formation of arches whose shape can be explained through a restricted random walk model (RRWM) [16,18,20]. Using this model, To *et al*. successfully reproduced the clogging probabilities at different outlet sizes. This model simply assumes a strict convex shape of the arch and equal probability for all allowable configurations. In reality, the presence of friction alters the strict convex shape of arch [21]. Additionally, this model only yields a geometric understanding of the arch structure without explanation of its dynamical origin. To understand the frequency of a clogging event and the avalanche statistics, Zuriguel *et al*. [15,22] further proposed an empirical Poisson process model. In this model, they introduce a single particle probability $p$ for each particle flowing out the orifice without forming an arch. The exponential distribution of avalanche sizes $s$ can be obtained by $P(s) = p^s(1-p)$. Nevertheless, it is difficult to reconcile this single particle model with the cooperative nature of the arch forming process as suggested by To *et al*. [18]. Later, Thomas *et al*. [23] suggest that clogging happens by the occurrence of certain multiple-particle clogging configurations, the clogging probability is determined by the ratio between these clogging configurations with their full configuration space. This model successfully explains the exponential distribution of avalanche sizes, but the nature of the configuration space and the multi-particle configuration remain elusive. Moreover, it is found that the flow speed can also strongly affect clogging [26]. An innovative silo discharge setup has been introduced with the flow rate controlled by a conveyor belt [24]. It is found that clogging probability can be empirically decomposed into two independent terms with one depends on outlet geometry and the other by particle kinematics. To develop a comprehensive understanding of the clogging phenomenon, it is necessary to incorporate the influence of all contributing factors, including the arch geometric structure, the clogging probability of a multiple-particle configuration, and the flow speed.

In this letter, we experimentally investigated the clogging process of grains in a two-dimensional hopper, and reveal that clogging happens due to the formation of some precursory chain structures in the flow. These chains induce clogging when they are mechanically stable enough to withstand the stress imposed from above. We further investigate the geometric structure and dynamic stability of these chains at different outlet sizes and flow speeds. And we find that clogging can be decomposed into two distinctive processes: chain formation, governed by outlet size and the particle contact geometrical structures, and its subsequent transformation into an arch, controlled by flow speed. Finally, we present an analytic expression for the clogging probability, which accurately reproduces our experimental results.

Our experimental setup [25] consists of a quasi-two-dimensional hopper with dimension of $100 \text{ cm}(H) \times 40 \text{ cm}(W)$, containing a monolayer of bi-disperse acrylic glass disks with diameters of 8 mm and 11.2 mm [Fig. 1(a)]. The average particle diameter, denoted as $d$, is 9.6 mm. The hopper outlet features two movable aluminum bars inclined at 45° from the horizontal direction, allowing continuously adjustment of the outlet width $D$. For our experiment, $D$ ranges from $2.5d$ to $4.8d$, and we repeat approximately 700 experimental realizations for each outlet width. To capture the full dynamic process between two clogging events, we employ a high-speed camera (Photron FASTCAM SA5) that records images at a rate of 500 frames per second.

Following the image processing procedures and tracking algorithms of our previous study [25-27], we can obtain the centroid coordinates and trajectories of disks with an uncertainty less than $3.1\times10^{-3}d$.

Clogging does not occur at regular intervals over time, with the discharge flux between two clogging events demonstrating avalanche-like statistics. Figure 1(b) shows the complementary cumulative distribution of the avalanche size *s*, obtained from a large number of clogging events. This distribution shows exponential behavior for all outlet widths *D*. Since the occurrence of clogging event is accompanied by an arch forming above the outlet, the clogging probability $P_c = 1/\langle s \rangle$ can be defined as the probability of forming an arch after an average outflow of $\langle s \rangle$ particles. Figure 1(c) shows that $P_c$ decreases quickly with the increase of *D/d*. This finding is consistent with previous works where $P_c$ has been fitted with various forms [5,7,15,28].

From a microscopic perspective, analyzing the images of clogged arches allow us to extract important geometrical features, including span, the number of particles involved and so on [29]. Similar to To's model [18], we characterize the geometry of these arch structures by examining the angle $\theta_i$, $i = 1, ..., n-1$ [the angle between the centroid of the $i_{th}$ particle and the $(i+1)_{th}$ particle with respect to the horizontal direction, inset of Fig. 2(a)]. We plot the probability distribution functions (PDFs) of $\theta$ for arches in Fig. 2(a). In our case, $\theta_i$ does not satisfy the strict decreasing trend as the RRWM by To *et al.* [18]. Instead, the presence of friction can lead to some local "defects" [21,30]. Therefore, the geometric rule of an arch is now $\theta_i + \delta\theta > \theta_{i+1}$, $\delta\theta = 20°$ in our case [25]. Additionally, we find $\theta_i > \theta_{i+2}$ [21], which ensures that an arch is globally convex to be mechanically stable. We note that these rules are mathematically complete in defining arch structure as the correlation of $\theta$ only extends to the next nearest neighbor as found in experiment. Equivalently, the arch structure can be described by $\phi_i$ (the angle spanned between a particle with two neighbors), which satisfies $\phi_i = \theta_i - \theta_{i-1} + 180°$ and $\phi_1 = \theta_1 + 135°$, $\phi_n = 135° - \theta_{n-1}$ [inset of Fig. 2(a)]. The corresponding two rules are now transformed into $\phi_i < 180° + \delta\theta = 200°$ and $\alpha_i = (\phi_i + \phi_{i+1})/2 \leq 180°$. Above two rules combined, which encompass more that 90% of the arch structures, set the geometric boundary for allowable arch configurations without specifying their probabilities.

To understand the nature of clogging and the associated phenomena in our experiment, we note that prior to a clogging event, some precursory arch-like structures already exist in the flow, which we denote as *chain*. These chains demonstrate a certain degree of mechanical rigidity, and continuously form and disintegrate during the flow process [inset of Fig. 1(a)]. Notably, when both ends of a chain come into contact with the hopper boundary, and if its stability is sufficient to withstand the flow impact, it transforms into an arch, ultimately leading to the arrest of the flow. By examining the occurrence and transformation of these chain structures, it is therefore possible to gain insights into the dynamics of arch formation and the underlying mechanisms of clogging. In this context, the probability of forming an arch $P_c$ can be mathematically expressed as:

$$P_c = \int p_c(v) \cdot P(v) dv = \int p_{chain} \cdot p_{a/c}(v) \cdot P(v) dv, \qquad (1)$$

where $P(v)$ is the PDF of the flow speed $v$ (see Supplemental Materials [25] for more details) and $p_c(v)$ is the conditional clogging probability at flow speed $v$. It can be further decomposed as the product of two terms: $p_{chain}$ is the chain-forming probability, which turns out to be independent of $v$. $p_{a/c}(v)$ is the probability of a chain transforming into an arch. This decomposition allows us to separate the clogging process into two distinctive mechanisms as we found that the first term $p_{chain}$ relies solely on the outlet shape and the particle contact geometry, while $p_{a/c}(v)$ only depends on the flow speed.

To obtain the chain-forming probability $p_{chain}$, we first define the chain structures [Fig. 1(a)]. We define a chain as a linear contacting structure whose two ends making contacts with the hopper boundaries, and the contacting particles between two ends satisfy the geometric rules specified for an arch structure, *i.e.*, $\theta_i + \delta\theta > \theta_{i+1}$, $\theta_i > \theta_{i+2}$. These rules are strong enough to exclude many mechanically unstable linear contacting structures which do not contribute to the arching process. We plot the PDFs of $\phi$ and $\alpha$ for the identified chains and arches in Fig. 2(b). Although $P(\phi)$ of the chains follow a Gaussian distribution with a most probable value about 165° which is similar as $P(\phi)$ of the arches, its distribution is significantly wider at large $\phi$. It suggests that a chain structure proximate to mechanical instability is less likely to transform into an arch.

Upon identifying different chain configurations, we can obtain $p_{chain}$ by calculating the ratio of the number of chains to the particles that flow through the outlet. It turns out it is determined by both particle contact geometry and outlet truncation effect. The probability of forming a *n*-particle chain at different outlet widths are shown in Fig. 3(a). The exponential decay observed at large *n* suggests that the probability of finding a chain with one additional particle remains constant [31]. We can explain the origin of this constant by looking at the geometric structures of particle contacts. When considering chain configurations, the fact that particles cannot overlap and the presence of bidispersity will impose a lower limit constraint on $\phi$ of 57°. At the same time, the upper limit constraint of 200° is stipulated by the definition of chain. Due to these constraints, the probability of extending a chain by one more particle is therefore: $\frac{1}{200-57}\int_{57}^{200} P(\phi)d\phi = e^{-1.03}$, where the normalization factor includes all possible particle contact configurations. The experimentally observed exponential decay can indeed be well-fit by $g(n) = g_0 e^{-1.03n}$ [the dash line in Fig. 3(a)], where $g_0 = 27.4$. The sudden decrease of the probability for small-$n$ configurations is due to the truncation of the outlet width since a chain must make contacts with both boundaries. To gain a simple theoretical understanding of this truncation effect, we suggest that the most probable chain configuration that $n$ particles can form is part of a regular polygonal structure. In this configuration, $\phi$ of each interior particle is the same, and the base angles on the left and right sides are 45° [inset of Fig. 3(b)]. Therefore, $\phi=180-90/n$. $\phi$ ranges from 162° to 167° for $n=5\sim7$, which is consistent with the experimentally observed mean value of $\phi$ being around 165° since $n=5\sim7$ chains are most common at different $D$. Real chains fluctuate around this mean structure, with the standard deviation of $\theta_i$ being $\sigma_{\theta_i}=0.4953$ (28.3°) [Fig. 2(a)]. Subsequently, the PDF $a_n(x)$ for a chain with $n$ particles having a horizontal projection length $x$ can be approximated by a Gaussian distribution with average value $x_n = \sum_{i=1}^{n-1} \cos\theta_i$ and standard deviation $\sigma_{x_n}^2 = \sum_{i=1}^{n-1} \sin^2\theta_i \cdot \sigma_{\theta_i}^2$. This simple model accurately captures the behavior of experimentally observed $a_n(x)$ at $D=3.49d$, see Fig. 3(b). Using this model, the chain-forming probability with its projection length greater than the outlet $D/d - \sqrt{2}/2$ ($\sqrt{2}/2$ comes from the contribution of two boundary particles) is $p_{chain} = \sum_{n=4}^{\infty} g(n) \int_{D/d-\sqrt{2}/2}^{\infty} a_n(x)dx$. As shown in the inset of Fig. 3(a), the theoretically calculated $p_{chain}$ are consistent with the experimental results.

To better understand why not all chains can transform into arches, it is important to recognize that chains are dynamical structures. Clogging can only happen when a chain maintains the geometric constraints specified for arches longer than a stopping time required to halt the flow. This help explain why chains with large $\phi$ and $\alpha$ are less likely to form arches as they are more likely to evolve across the geometric constraints within this period. We experimentally confirm that 40% of the chains break at the particle with $\phi \sim 200°$, 49% at the location with $\alpha \sim 180°$, and 10% where both $\phi \sim 200°$ and $\alpha \sim 180°$.

In Fig. 4(b), we plot the experimentally observed overall probability of chains with different number of particles transforming into the arch as a function of flow speed $v$ at different $D$. Interestingly, when we take $n$th root of $p_{a/c}(v,n)$ for chain of length $n$ at different $D$, denoted as $p_s(v)$, they collapse onto the same curve. Specifically, $p_s(v)$ decreases monotonically with increasing $v$, and asymptotes to a non-unity value as $v$ approaches zero. This universal behavior of $p_s(v)$, regardless of the chain structure and outlet width, suggests that it is a quantity which only depends on $v$ and can be analyzed on the single particle level.

To understand this behavior, it is important to analyze how the chain structure evolves over time. The mean-squared displacement (MSD) of $\phi$ associated with each particle is:

$$\left\langle \phi^2(\Delta t) \right\rangle = \frac{1}{N} \sum_{j=1}^{j=N} \left| \phi_j(\Delta t) - \phi_j(0) \right|^2 , \qquad (2)$$

where $N$ averages over all possible chains, irrespective of the number of particles they consist of. Figure 4(a) shows that $\left\langle \phi^2(\Delta t) \right\rangle \sim \Delta t^2$, which suggests that $\phi$ evolves linearly as a function of $\Delta t$ on short time scales, with a speed $\dot{\phi}$. This contrasts with the sub-diffusive behavior of $\phi$ when a stagnant arch is vibrated [32,33]. To understand what determines $\dot{\phi}$, we found that at a specific $\phi$, the variance $\sigma(\dot{\phi})\big|_{\phi=const}$ of $\dot{\phi}$ satisfies a linear relation with $v$: $\sigma(\dot{\phi})=0.14v+1.24$, see Fig. 4(c). Furthermore, by normalizing $\dot{\phi}$ with $\sigma(\dot{\phi})\big|_{\phi=const}$, the profiles of $\dot{\phi}$ and $\phi$ for different outlet widths can be collapsed onto a single curve

$\bar{\phi}/\sigma(\dot{\phi})=0.025\phi-4$, see Fig. 4(d). This behavior is because that $\phi$ and $\dot{\phi}$ for all outlet widths satisfy a universal joint gaussian distribution with a covariance dependent on $\phi$:

$$P(\dot{\phi},\phi) = \frac{1}{\sqrt{2\pi}\sigma(\dot{\phi})}\exp\left[-(\dot{\phi}-\bar{\dot{\phi}})^2/2\sigma(\dot{\phi})^2\right] \times \frac{1}{\sqrt{2\pi}\sigma(\phi)}\exp\left[-(\phi-\bar{\phi})^2/2\sigma(\phi)^2\right]. \quad (3)$$

The non-zero value of $\sigma(\dot{\phi})$ as $v$ tends to zero implies that even under quasi-static flow condition, the chain structure can still undergo structural rearrangement. This is consistent with the observation in Ref. [34].

Based on $P(\dot{\phi},\phi)$, we can theoretically calculate the probability of a chain transforming into an arch on the single particle level. We start with the configuration of a single particle and its nearest neighbor constraint within a chain, and calculate its probability of maintaining an arch configuration after the flow stopping time $\delta t$ by the following integral:

$$p_s(v) = \frac{1}{A}\int_{\phi_{1lower}}^{\phi_{1upper}}\int_{\dot{\phi}_{1lower}}^{\dot{\phi}_{1upper}}\int_{\phi_{0lower}}^{\phi_{0upper}} P(\phi_0)P(\dot{\phi}_1,\phi_1)\mathrm{d}\phi_0\mathrm{d}\dot{\phi}_1\mathrm{d}\phi_1. \quad (4)$$

Where $A$ is the normalization factor. The details regarding the integral can be found in the Supplemental Materials [25]. We calculate this quadruple integral numerically since it is difficult to reach an analytical solution. The numerical results exhibit excellent agreement with the experimental observed $p_s(v)$, as shown in Fig. 4(b). For a chain consisting of $n$ particles, the probability that a chain can transform into an arch is therefore $p_s(v)^n$.

Finally, the clogging probability $P_c$ can then be calculated as:

$$P_c = \int_v [\sum_n g(n) \int_{D/d-\sqrt{2}/2}^{\infty} a_n(x)dx \cdot p_s(v)^n] \cdot P(v)dv. \quad (5)$$

In the inset of Fig. 1(c), the theoretically obtained $P_c$ based on Eq. (5), which incorporates the influence of outlet size, friction, and flow speed, agrees very well with the experimentally measured results. Furthermore, Eq. (5) yields the theoretical results of the PDF of number of particles *n* that contained in the arch, which also agree nicely with the experimental results (see Supplemental Materials [25]).

In conclusion, we have achieved a microscopic understanding of the physical mechanism for the clogging phenomenon in gravity-driven granular hopper flow. By considering the formation and mechanical stability of specific precursory chain structures during flow, we are able to calculate the overall clogging probability, which demonstrates good agreement with experimental results. These findings suggest that clogging in principle can still occur for large outlet size, despite the significant increase of the discharge size as outlet size increases.

The work is supported by the National Natural Science Foundation of China (No. 11974240, No. 12274292), the Science and Technology Innovation Foundation of Shanghai Jiao Tong University (No. 21X010200829).

Corresponding author

*yujiewang@sjtu.edu.cn

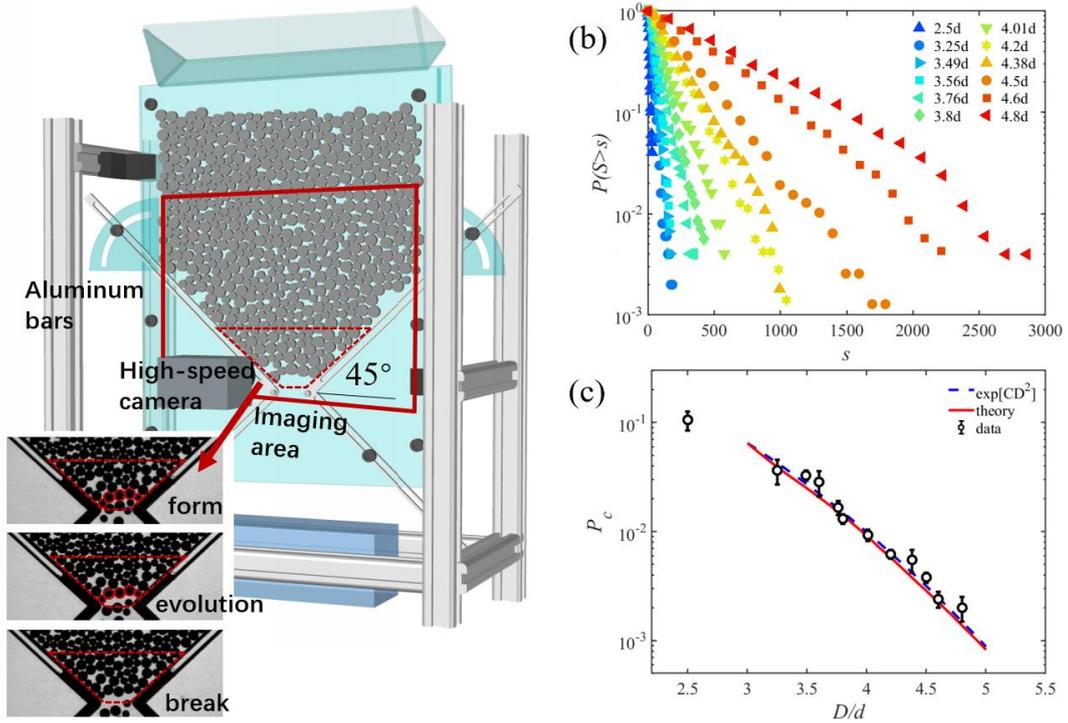

FIG. 1. (a) Schematic of the experimental setup. The high-speed camera images the area within the red frame and chain structures are analyzed in the red dashed trapezoidal regime. Lower panel: raw images for the formation, evolution and breaking processes of a 5-particle chain (marked with red circles) at $D=3.49d$. (b) Complementary cumulative distribution of the avalanche size $s$ for different $D$. (c) Experimentally obtained clogging probability $P_c$ (black dots) as a function of $D/d$ and the corresponding theoretical results from Eq. (5) (red line). The dash curve is the empirical fitting using $P_c = \exp[CD^2]$.

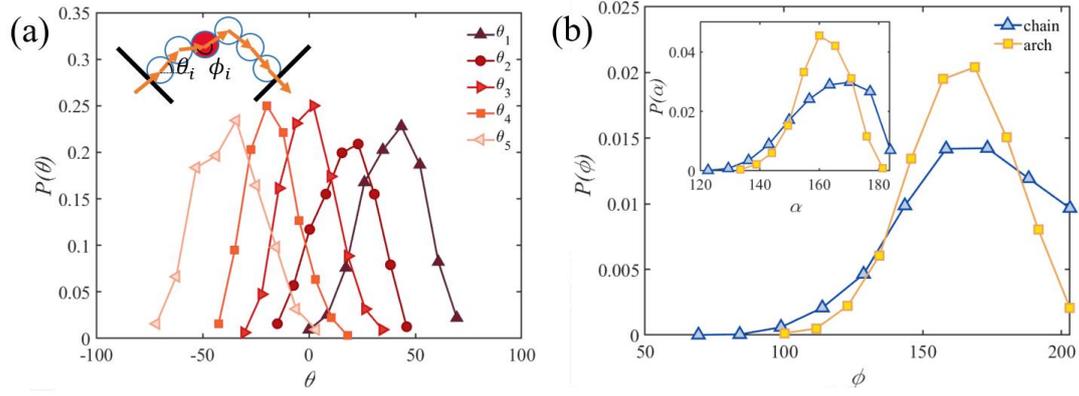

FIG. 2. (a) PDFs of $\theta_i$ (from right to the left, $i=1, 2, \ldots, 5$) for arches consisting of 6 particles. Inset: schematic diagram of a chain structure. (b) PDFs of $\phi_i$ for chains (triangle) and arches (square) at $D = 3.49d$. Inset: PDFs of $\alpha_i$ for chains (triangle) and arches (square) at the same $D$.

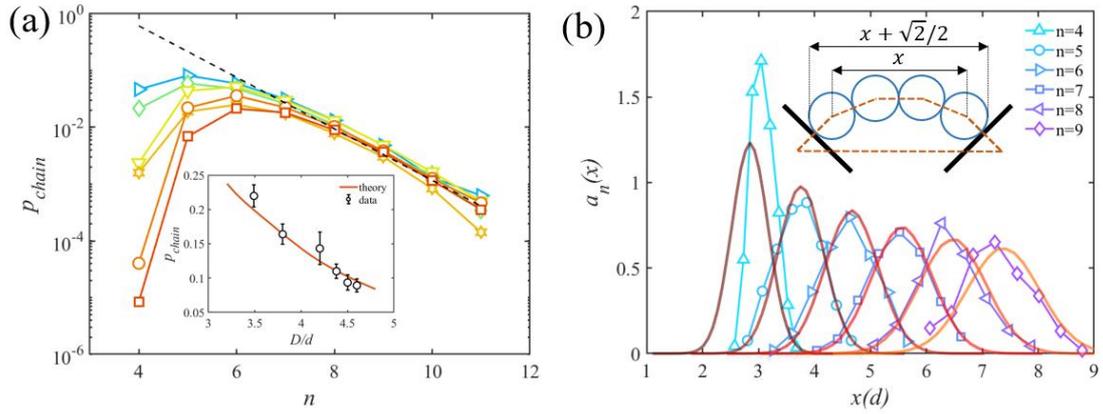

FIG. 3. (a) Chain-forming probability $p_{chain}$ for $n$-particle chains at different $D$. The dashed line is the empirical fitting of $g(n)$. Inset: experimentally obtained $p_{chain}$ (dots) and the corresponding theoretical prediction (curve) as a function of $D$. (b) Experimentally obtained PDFs $a_n(x)$ for a chain with $n$ particles to have a horizontal projection length $x$ ($n = 4$ to $n = 9$). Solid curves are results given by the polygon model. Inset: schematic diagram of the polygon model.

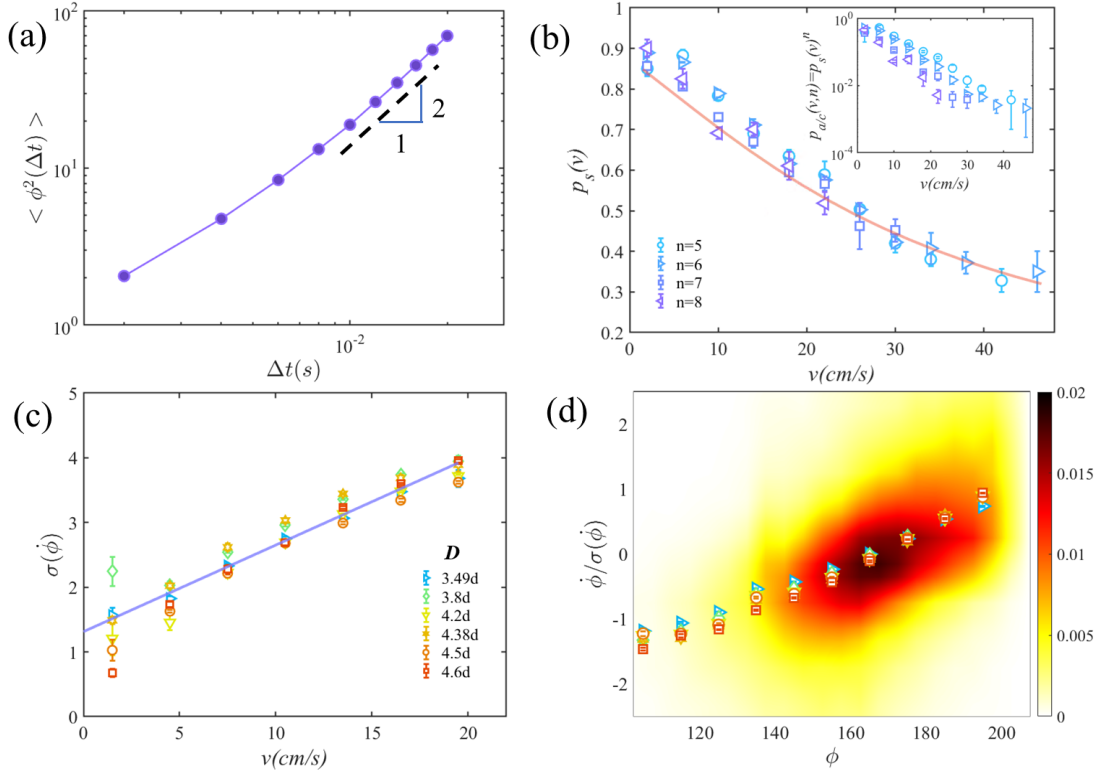

FIG. 4. (a) MSD for chain structures as a function of $\Delta t$ at $D = 3.49d$. The black dash line is a guide for eye with MSD $\sim \Delta t^2$. (b) Single-particle probability of a chain transforming into an arch $p_s(v)$ as a function of $v$ at $D=4.5d$. Solid curve is the result of the theoretical model. Inset: Multiple-particle probability of a chain transforming into an arch $p_{a/c}(v,n) = p_s(v)^n$ as a function of $v$ at $D=4.5d$. (c) Variance $\sigma(\dot{\phi})$ as a function of $v$, which satisfies the same linear relation for all $D$. (d) Universal joint probability density distribution $P(\dot{\phi}/\sigma(\dot{\phi}),\phi)$ for all $D$. The relation between the rescaled $\bar{\dot{\phi}}$ and $\phi$ can be collapsed onto a single curve for all $D$.

# Supplemental Materials for

# Experimental Study of Granular Clogging in a Two-Dimensional Hopper

Shuyang Zhang,[1] Zhikun Zeng,[1] Houfei Yuan,[1] Zhifeng Li,[4] and Yujie Wang[1,2,3,*]

[1]*School of Physics and Astronomy, Shanghai Jiao Tong University, Shanghai 200240, China*
[2]*State Key Laboratory of Geohazard Prevention and Geoenvironment Protection, Chengdu University of Technology, Chengdu 610059, China*
[3]*Department of Physics, College of Mathematics and Physics, Chengdu University of Technology, Chengdu 610059, China*
[4]*School of Mathematics and Physics, University of South China, Hengyang 421001, China*

Corresponding author

[*]yujiewang@sjtu.edu.cn

## 1. Experimental device

Our experimental setup consists of a quasi-two-dimensional hopper with dimension of $100 \text{ cm}(H) \times 40 \text{ cm}(W)$, constructed using two transparent acrylic glass plates [Fig. 1(a) of the main text]. These plates are separated by two aluminum strips along the boundary, creating a 6 mm gap between them. To form the hopper outlet, two movable aluminum bars are used, which are inclined at an angle of with respect to the side walls. The outlet width, denoted as $D$, can be continuously adjusted by raising or lowering the side bars. For our experiment, $D$ ranges from $2.5d$ to $4.8d$. To ensure that the particle flow is two-dimensional, we use bi-disperse disks made of acrylic glass with a thickness of 5 mm and diameters of 8 mm and 11.2 mm. The average particle diameter, denoted as $d$, is 9.6 mm. These disks are laser-cut from smooth acrylic glass plates, minimizing friction interactions with the front and back plates. Prior to being fed into the hopper, the particles are uniformly mixed with a number ratio of 2:1 to achieve roughly equal area fractions. When a clogging event happens, we then use a rod to lightly poke the arch at the outlet to restart the flow. The rod is quickly removed to avoid interfering with the flow. To capture the full dynamic process between two clogging events, we employ a high-speed camera (Photron FASTCAM SA5) that records images at a rate of 500 frames per second. The imaging area is about $30\times30$ cm$^2$ which covers the outlet and the region above it, and only disks within a $4d$ distance from the outlet are analyzed in our experiment since experimentally no arch can form above this height, see Fig. 1(a) of the main text. For better statistics, the clogging process is repeated approximately 700 times for each outlet width.

**2. Image processing and particle tracking algorithm**

After obtaining the flow image sequence near the outlet of the hopper flow, we can obtain the coordinates and size of each particle with an error less than $3.1\times10^{-3}d$ by following similar image processing techniques as described in the previous work [26,27], as shown in Fig. S1. Subsequently, the particle trajectories can be determined using a tracking algorithm: a particle in the first frame is identified as the same particle in the second frame when they have the smallest spatial distance. This tracking algorithm has been proven effective when the particle displacements are less than $1/2d$ between two successive frames. In our experiment, the particle displacements are less than $1/4d$ in practice.

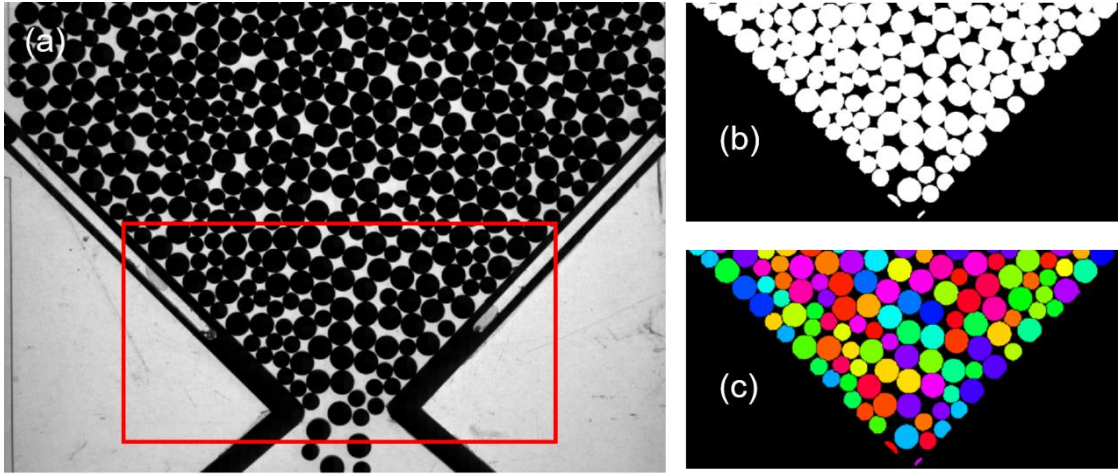

FIG. S1. (a) Raw image of the granular hopper flow. (b) Binarized image of (a), shown within the red frame. (c) Segmented image using a watershed algorithm.

### 3. Flow speed

The instantaneous flow speed $v$ at a specific time $t$ is calculated as $v = \frac{1}{m}\sum_{i=1}^{m}(v_{ix}^{2} + v_{iy}^{2})^{1/2}$, where $v_{ix}$ and $v_{iy}$ represent the horizontal and vertical displacements of particle $i$ between two consecutive frames, respectively, and the average is taken over all grains in the vicinity of the outlet region. For different outlet widths, we obtain the probability distribution function (PDF) of $v$, $P(v)$, by calculating the flow speed $v$ for all experimental realizations, as shown in Fig. S2(b). Subsequently, we calculate the average flow speed $\bar{v}$ at an outlet width by averaging over the whole flow process between two clogging events and all experimental realizations.

We find that for $D/d$ ranging from 3 to 5, the average flow speed $\bar{v}$ shows a linear dependence on the outlet width $D$, as shown in Fig. S2(a). Furthermore, $P(v)$ can be reasonably approximated by a Gaussian distribution $P(v) = \frac{1}{\sqrt{2\pi}\sigma_v}\exp\left[\frac{-(v-\bar{v})^2}{2\sigma_v^2}\right]$ with a variance $\sigma_v = 7$ that applies to all $D$.

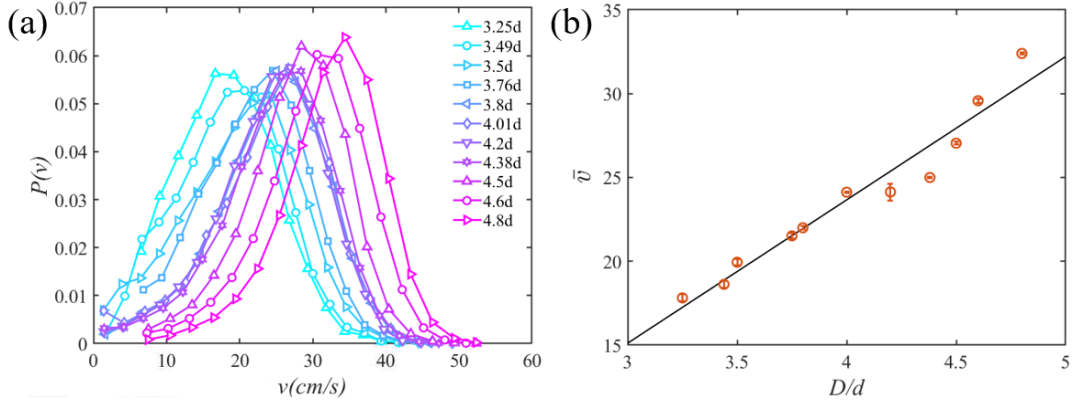

FIG. S2. (a) PDFs of flow speed $v$ at different $D$. (b) Averaged flow speed $\bar{v}$ as a function of $D/d$. The solid line represents the linear fitting of $\bar{v}=8.5D/d-10$.

**4. Defects in the restricted random walk model**

Due to the influence of friction, $\theta_i$ of a particle and $\theta_{i+1}$ of its previous neighbor in an arch does not strictly follow the decreasing trend $\theta_i > \theta_{i+1}$. The presence of local "defects" result in $\delta\theta = \theta_{i+1} - \theta_i > 0$. The PDF of $\delta\theta$ for real arch structures at different $D$ is shown in Fig. S3. To impose a modified geometric rule that could include most of the arch configurations, we set $\delta\theta = 20°$ and modify the geometric rule to $\theta_i + 20° > \theta_{i+1}$. This modification ensures that more than 90% of the arches can now be included.

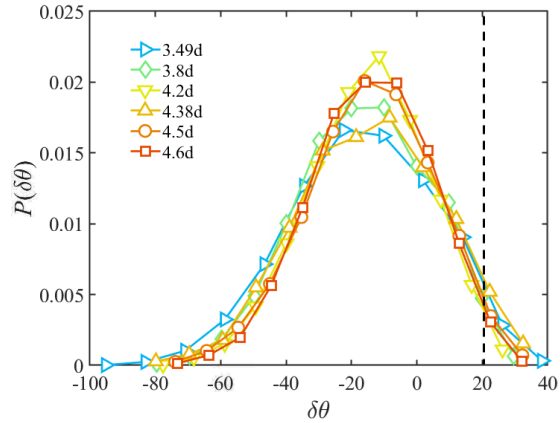

FIG. S3. PDF of $\delta\theta$ at different $D$. The black dash line represents the threshold at $\delta\theta = 20°$.

## 5. The integral calculation of $p_s(v)$

Here we provide a detailed description of the procedure for numerically calculating the integral from which we can obtain the single-particle probability $p_s(v)$ of a chain transforming into an arch. Firstly, we randomly sample $\phi_1 \in [100°, 200°]$ of grain 1 from $P(\phi)$, along with its initial angular speed $\dot{\phi}_1$ from $P(\dot{\phi})$. Since $\phi$ evolves linearly in time, $\phi_1$ evolves to $\phi_1' = \phi_1 + \dot{\phi}_1 \delta t$ after a characteristic stopping time $\delta t$. Thus we can determine the lower and upper limits of $\dot{\phi}_1$, $\dot{\phi}_{1lower} = (100 - \phi_1)/\delta t$ and $\dot{\phi}_{1upper} = (200 - \phi_1)/\delta t$, as the chain should maintain an arch configuration $\phi_1' \in [100°, 200°]$. In addition, to satisfy the nearest neighbor constraint in defining a chain, $\alpha' = (\phi_0 + \phi_1')/2$ need in $[125°, 180°]$, where $\phi_0$ is the angle of the neighboring grain which also satisfies $P(\phi)$. This implies that the lower and upper limits for $\phi_0$ is $\phi_{0lower} = \max(100, 250 - \phi_1')$ and $\phi_{0upper} = \min(200, 360 - \phi_1')$. When $\phi_1'$ is fixed, the probability that $\phi_0$ is within these limits is:

$$p_0(\phi_1') = \int_{\phi_{0lower}}^{\phi_{0upper}} P(\phi_0) d\phi_0 .$$

Eventually, the single-particle probability $p_s(v)$ of a chain transforming into an arch can be calculated as:

$$p_s(v) = \frac{1}{A} \int_{\phi_{1lower}}^{\phi_{1upper}} \int_{\dot{\phi}_{1lower}}^{\dot{\phi}_{1upper}} \int_{\phi_{0lower}}^{\phi_{0upper}} P(\phi_0) P(\dot{\phi}_1, \phi_1) d\phi_0 d\dot{\phi}_1 d\phi_1 ,$$

where $A = \int_{\phi_{0lower}}^{\phi_{0upper}} P(\phi_0) d\phi_0 \cdot \int_{\phi_{1lower}}^{\phi_{1upper}} P(\phi_1) d\phi_1$ is the normalization factor because the chain structures are defined when $\phi_{0,1} \in [100°, 200°]$ and $\alpha \in [125°, 180°]$. The stopping time $\delta t = 0.018$ s here is an empirically fitted value which enables the theoretical results to best match experimental observations for different outlet widths. Notably, the numerical results derived from the theoretical model demonstrate remarkable agreement with the experimentally observed $p_s(v)$, as presented in Fig. 4(d) of the main text.

## 6. PDFs of the number of particles *n* that contained in the arch

From Eq. (5), we can obtain the PDF for the number of particles *n* that form the arch at different outlet widths as:

$$P(n) = P_c|_{n=n_i}/P_c = \frac{1}{P_c}\left(g(n)\int_{D/d-1}^{\infty} a_n(x)dx\right)\int_v p_s(v)^n \cdot p(v)dv,$$

where $n_i$ ranges from 4 to 11. Figure S4 demonstrates that good agreement exists between $p_{n_i}$ predicted by our theoretical model and the experimental ones.

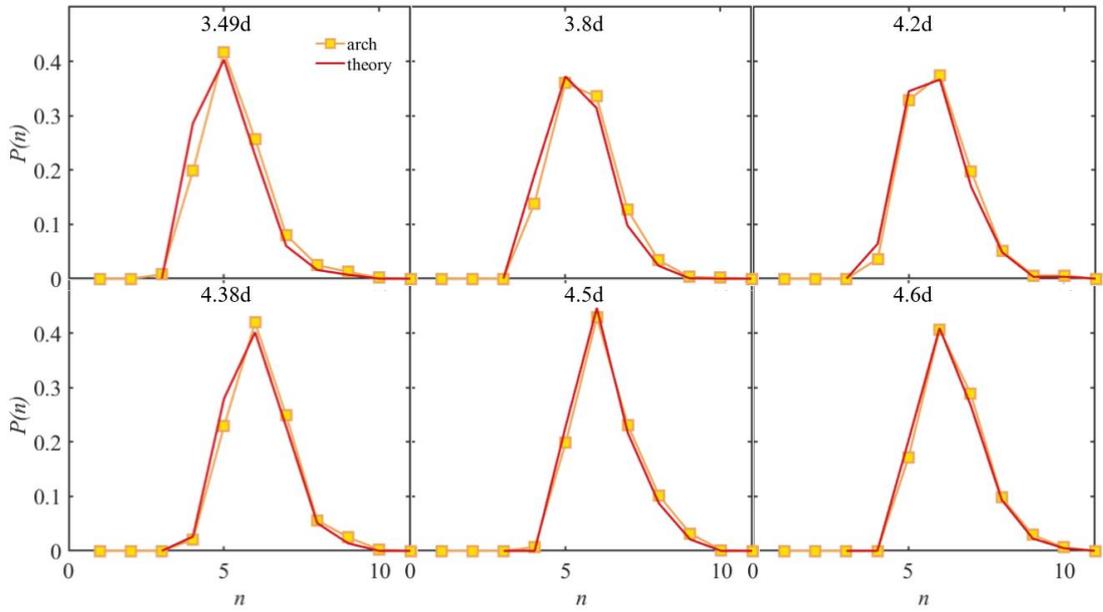

FIG. S4. PDFs of the number of particles *n* that form the arch for different *D*. Red curves and yellow marks are results given by the theoretical model and the experimental measurement respectively.